\documentclass[a4paper, amsfonts, amssymb, amsmath, reprint, showkeys, nofootinbib, twoside]{revtex4-1}
\pdfoutput=1
\usepackage[english]{babel}
\usepackage{graphicx}
\usepackage[colorinlistoftodos, color=green!40, prependcaption]{todonotes}
\usepackage[pdftex, pdftitle={Article}, pdfauthor={Author}]{hyperref} 
\begin{document}
\title{Excitation and voltage-gated modulation of single-mode dynamics in a planar nano-gap spin Hall nano-oscillator}

\author{Lina Chen$^{1,2}$}
\author{Yu Chen$^1$}
\author{Zhenyu Gao$^2$}
\author{Kaiyuan Zhou$^2$}
\author{Zui Tao$^2$}
\author{Yong Pu$^1$}
\email{puyong@njupt.edu.cn}
\author{Tiejun Zhou$^3$}
    \email{tizhou@hdu.edu.cn}
\author{Ronghua Liu$^{2}$}
    \email{rhliu@nju.edu.cn}
    \affiliation{
$^1$School of Science, Nanjing University of Posts and Telecommunications, Nanjing 210023, China\\
$^2$Jiangsu Provincial Key Laboratory for Nanotechnology, School of Physics\\ and National Laboratory of Solid State Microstructures, Nanjing University, Nanjing 210093, China\\
$^3$Centre for Integrated Spintronic Devices, School of Electronics and Information, Hangzhou Dianzi University, Hangzhou 310018, China}


\begin{abstract}
We experimentally study the dynamical modes excited by current-induced spin-orbit torque and its electrostatic gating effect in a 3-terminal planar nano-gap spin Hall nano-oscillator (SHNO) with a moderate interfacial perpendicular magnetic anisotropy (IPMA). Both quasilinear propagating spin-wave and localized "bullet" modes are achieved and controlled by varying the applied in-plane magnetic field and driving current. The minimum linewidth shows a linear dependence on the actual temperature of the active area, confirming single-mode dynamics based on the nonlinear theory of spin-torque nano-oscillation with a single mode. The observed electrostatic gating tuning oscillation frequency arises from voltage-controlled magnetic anisotropy and threshold current of SHNO via modification of the nonlinear damping and/or the interfacial spin-orbit coupling of the magnetic multilayer. In contrast to previously observed two-mode coexistence degrading the spectral purity in Py/Pt-based SHNOs with a negligible IPMA, a single coherent spin-wave mode with a low driven current can be achieved by selecting the ferromagnet layer with a suitable IPMA because the nonlinear mode coupling can be diminished by bringing in the PMA field to compensate the easy-plane shape anisotropy. Moreover, the simulations demonstrate that the experimentally observed current and gate-voltage modulation of auto-oscillation modes are also closely associated with the nonlinear damping and mode coupling, which are determined by the ellipticity of magnetization precession. The demonstrated nonlinear mode coupling mechanism and electrical control approach of spin-wave modes could provide the clue to facilitate the implementation of the mutual synchronization map for neuromorphic computing applications in SHNO array networks.
\end{abstract}


\maketitle

\section{Introduction}
Spin Hall nano-oscillator~\cite{demidov,liurh} is a new alternative to traditional spin-transfer-torque nano-oscillators (STNOs)~\cite{tsoi,kiselev,Bfang,kshi,kaiyuan} based on current-perpendicular-to-plane spin-valve or magnetic tunnel junction (MTJ) structures. SHNOs consist of a single ferromagnet (FM) and heavy metal (HM) with a strong spin-orbit coupling (SOC) bilayer,and utilize bulk spin Hall effect (SHE) of HM~\cite{SHE} and interfacial Rashba-Edelstein effect (IREE) at the HM/FM interface~\cite{IREE} to generate an out-of-plane spin current under passing an in-plane electric-current through the bilayer plane~\cite{yanglp,wang2022}. Thanks to this simple in-plane structure for easy fabrication process and flexible and scalable two-dimensional architecture, SHNO constructed of numerous materials or different geometries, including ferromagnet metals and insulators with in-plane and out-of-plane magnetization, have been recently intensity studied~\cite{zholud,Nano-SHNO,duan,liu2015,collet,jung,Fulara,liupra2020,lli}. However, previous reports have proved that the planar nanogap SHNO based on an extended Pt/Py bilayer with an easy-plane magnetization prefers simultaneous excitation of two dynamical modes, significantly degrading spectra purity of SHNO due to their mode-coupling~\cite{liurh,zholud,ulrichs,prb2019}. Meanwhile, SHNO with a suitable PMA extended film exhibits the dynamical bubble mode with the spectrum consisting of a primary peak and two sidebands at small in-plane magnetic fields, and mode transition from propagating mode to self-localized bullet mode at large in-plane magnetic fields~\cite{liu2015,liu2019}. The very recent study also revealed that nonlinear damping could be controlled by the ellipticity of magnetization precession which was determined by magnetic anisotropy of the device~\cite{nonlineardamping}. Therefore, it is essential to experimentally explore the spectral characteristics of SHNOs with a moderate PMA to facilitate their promised applications.

Additionally, synchronization with external rf source in individual SHNO~\cite{syn-shno} and as well as mutual synchronization between multiple coupled SHNOs in one-dimensional chains and two-dimensional arrays~\cite{awad,SHNO-2D} are drawing increasing attention because these mutually coupled nonlinear spin-based oscillators are promising to mimic human brain processing functions to develop new high-speed and low-power neuromorphic computing~\cite{SHNO-2D,vowel,Jiang}. SHNOs are well suited to neuromorphic applications due to their intrinsically nonlinear behavior and strongly nonlinear interaction between oscillators or external stimulation signals. Therefore, exploring a more energy-efficient approach to electrically turn individual ones and mode coupling in SHNOs network arrays is important before scaling the neuromorphic computing to large nonlinear dynamics neural networks for application to a wide range of complex, high-dimensional tasks. Previous works in the conventional spintronics field have proved that voltage-controlled magnetic anisotropy is a highly energy-efficient approach to control magnetization~\cite{matsukura}, e.g., magnetization reversal and procession, compared to the current-based approach. In addition, the in-plane configuration of SHNOs can easy to achieve both current- and voltage-based collaborative control of nonlinear dynamics in three-terminal SHNOs~\cite{liu2017,fulara2,choi}.

Since [Co/Ni] multilayer has a moderate interfacial PMA, low Gilbert damping, and large anisotropy magnetoresistance (AMR) as well as Permalloy (Ni$_{80}$Fe$_{20}$)~\cite{daalderop,rippard,PMA2022}, here we adopt 1.7 nm thick [Co/Ni]$_3$/Co multilayer as the ferromagnetic layer to built three-terminal SHNOs and experimentally study the effects of current and electrostatic gating on SOT-induced magnetization oscillation. In addition to quasilinear propagating spin-wave emerging at low in-plane fields and small currents, a single self-localized "bullet" spin-wave mode with a frequency below ferromagnetic resonance (FMR) frequency can also be excited at large in-plane fields and large currents. In contrast to Py/Pt-based SHNO, the two-mode coexistence-induced decoherence phenomenon is not observed in our thin [Ni/Co]/Pt-based SHNO with a moderate PMA. Our micromagnetic simulations reveal that the perpendicular magnetic anisotropy via diminution of nonlinear mode-coupling and nonlinear damping can significantly lower the threshold current and suppress the previously observed secondary spin wave mode localized near the two edges of the center bullet mode in nano-gap SHNOs. Furthermore, the three-terminal SHNO shows a 200 MHz frequency tunability (7 MHz/V) by voltage gating due to the electric field modulating the threshold current and IPMA.

\section{Experimental section}

\begin{figure}[htbp]
\centering
\includegraphics[width=0.45\textwidth]{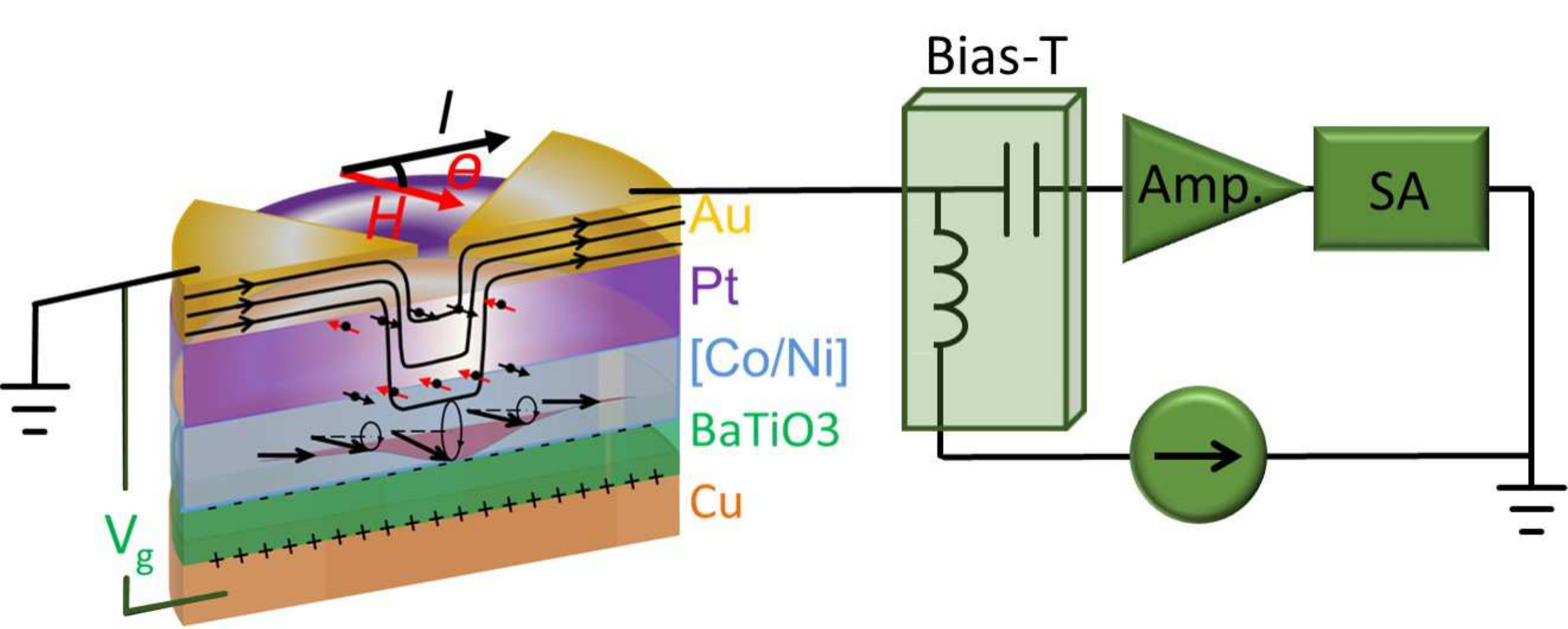}
\caption{The cross-sectional view of voltage-control SHNO device structure with multilayer order and the experimental setup with the directions of current flow $I$ and the applied magnetic field $H$, the angle $\theta$ and electrostatic gating $V_g$. The region of spin current-induced oscillating magnetization is localized in the multilayer [Co/Ni] under the central nanogap.}\label{fig1}
\end{figure}

Figure~\ref{fig1} shows the schematic of our test device structure and the experimental setup. Our device is based on a stacked multilayer Cu(30)/BaTiO$_3$(30)/[Co(0.2)/Ni(0.3)]$_3$/Co(0.2)/Pt(4) deposited on an annealed sapphire substrate at room temperature (RT). All thicknesses are given in nanometers. The [Co(0.2)/Ni(0.3)]$_3$/Co(0.2)/Pt(4) (abbreviated as [Co/Ni]$_3$/Co/Pt) multilayer disk with 4 $\mu$m diameter and its top two triangle-shaped Au electrodes with approximately 100 nm gap were electrically isolated from the 30 nm thick Cu bottom gating electrode by a 30 nm thick dielectric layer BaTiO$_3$ [Fig.~\ref{fig1}]. The 100 nm thick head-to-head triangular Au electrodes as two in-plane point contacts are used to inject current locally into the [Co/Ni]$_3$/Co/Pt multilayer disk and achieve the highly localized current density in the Pt layer within the gap area. The 30 nm thick dielectric layer BaTiO$_3$ is grown by using ultrahigh vacuum pulsed laser deposition with 80 mTorr oxidant background gas (99\% O$_2$ + 1\% O$_3$) at RT~\cite{BTO}. A KrF excimer laser ($\lambda$ = 248 nm) with a repetition rate of 3 Hz and a laser flounce of 1 J/cm$^2$ was used. The other metal layers are grown at RT by magnetron sputtering with base pressure less than 2 $\times$ 10$^{-8}$ Torr. The device was fabricated by a combination of magnetron sputtering and electron beam lithography. In this three terminals device, when the in-plane electrical current with a high current density ($\sim$ 10$^8$A/cm$^2$) passes through the Pt layer within the 100 nm wide nanogap, it will generate the spin currents due to the bulk SHE in Pt(2) layer and IREE at both Co/Pt and BaTiO$_3$/Co interfaces perpendicularly injected into the [Co/Ni]$_3$/Co multilayer. Similar to the previously studied nanogap SHNOs~\cite{liurh,liu2015}, all the measurements of microwave spectra described below are performed at in-plane magnetic field geometry with an in-plane angle $\theta$ between the in-plane field $H$ and the direction of electrical current $I$.

\section{Results and discussion}
\subsection{Dependence of spectral characteristics on current and magnetic field at $V_g$ = 0}

\begin{figure}[!t]
\centering
\includegraphics[width=0.45\textwidth]{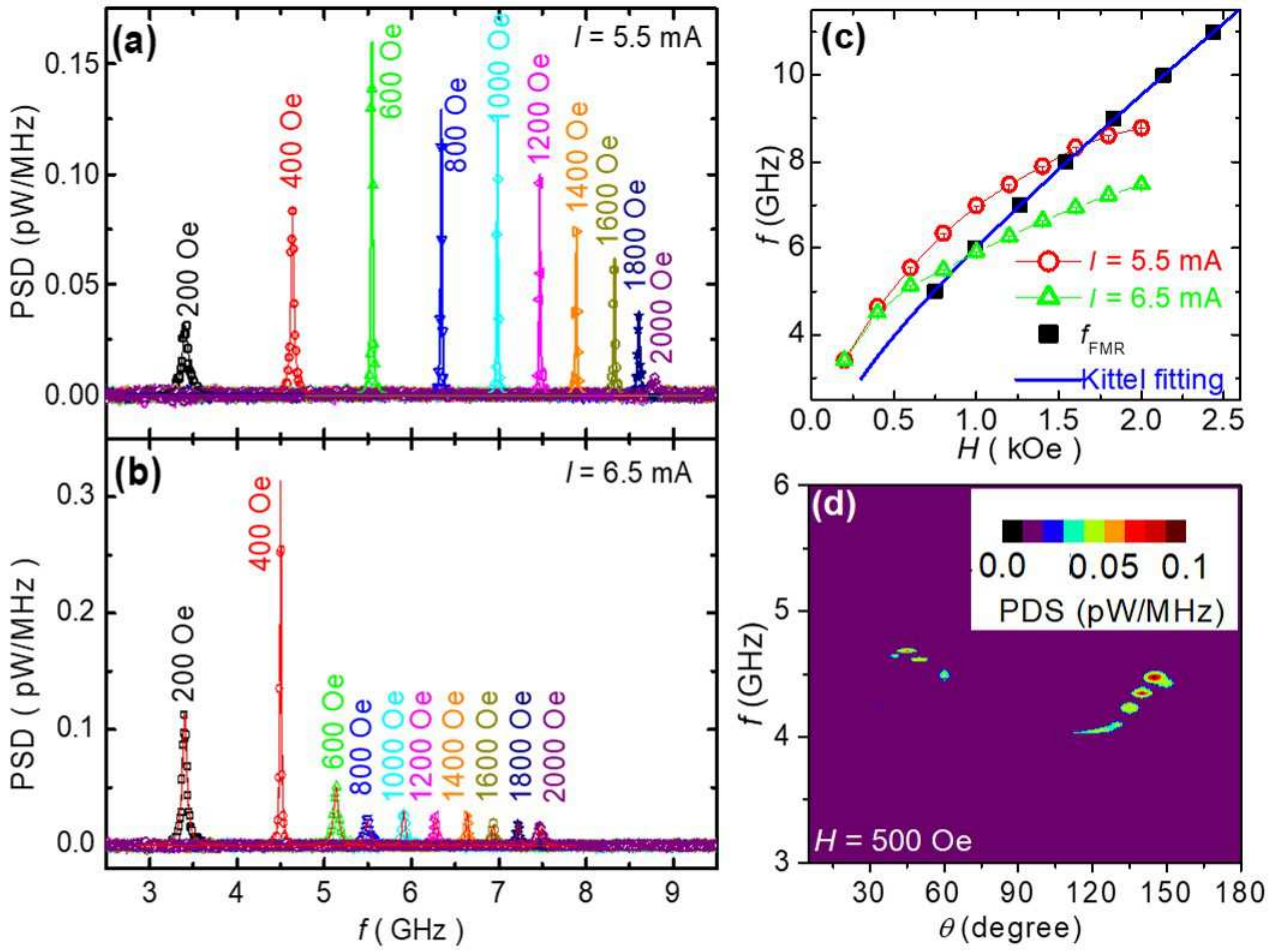}
\caption{Dependence of the microwave generation characteristics of SHNO on the magnitude of magnetic field $H$ and the angle $\theta$ between $H$ and current $I$ at 6 K. (a - b) Spectra obtained at $\theta$ = 120$^\circ$,  labeled $H$ and $I$ = 5.5 mA (a) or 6.5 mA (b). (c) Dependence of the uniform ferromagnetic resonance (FMR) frequency $f_{FMR}$ (solid squares) and the auto-oscillation frequency $f_{auto}$ (hollow symbols) obtained at 5.5 mA and 6.5 mA  on the applied magnetic field $H$, which were determined by spin-torque FMR (ST-FMR) technique and fitting the PSD spectra of (a) and (b) with Lorentzian function. The solid curve is the result of fitting the FMR data with the Kittel formula. (d) Pseudocolor maps of the dependence of the generated microwave spectra on the angle $\theta$ at $H$ = 500 Oe, $I$ = 7 mA.}\label{fig2}
\end{figure}

\begin{figure*}[!t]
\centering
\includegraphics[width=0.95\textwidth]{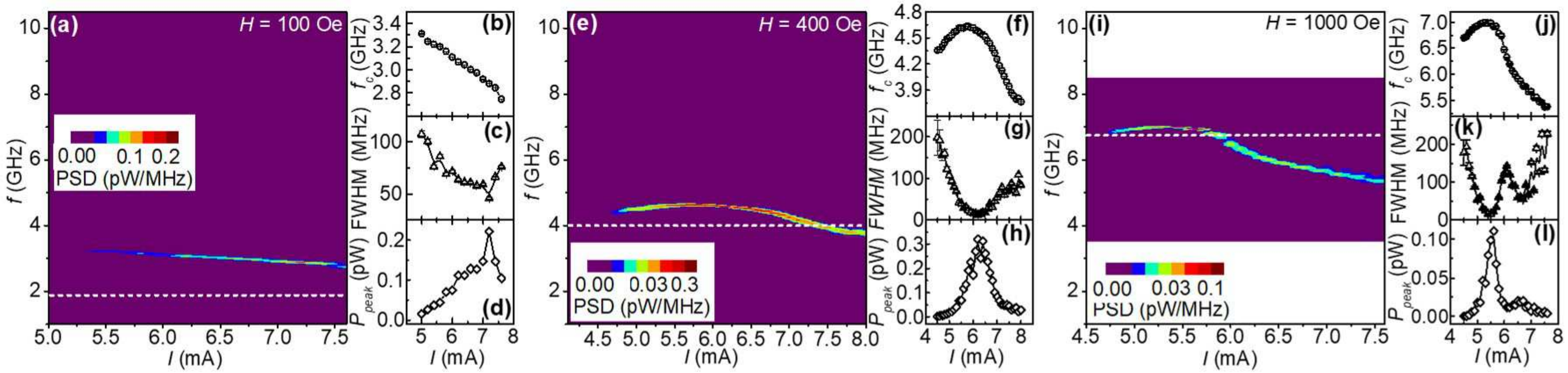}
\caption{Dependence of the microwave generation characteristics of SHNO on current at 6 K, $\theta$ = 120$^\circ$ and three different $H$. (a) Pseudocolor maps of the dependence of the generated microwave spectra obtained at $H$ = 100 Oe on current. (b)-(d) Dependence of the central generation frequency $f_c$ (b), the full width at half maximum (FWHM) (c) and the intensity peak $P_{peak}$ (d) on current $I$ were determined by fitting the power spectra of (a) with Lorentzian function. (e-h) and (i-l) Same as (a-d), at $H$ = 400 Oe and 1000 Oe, respectively. The dotted lines represent the corresponding FMR frequencies $f_{FMR}$ of the device.
}\label{fig3}
\end{figure*}

To obtain better microwave spectra by suppressing thermal fluctuation broadening, we performed the spectra measurements at a cryogenic temperature $T$ = 6 K. Spin current-induced auto-oscillations, indicated by the abrupt emergence of a sharp peak in the microwave spectra, can be achieved above the onset current $I_{on}$ $\sim$ 4.5 - 5.0 mA in the studied fields of 0.1 kOe to 2 kOe. The value of $I_{on}$ is smaller than 5.7-6.1 mA in Py(3)/Pt(2)-based SHNO~\cite{liu2017}, suggesting more energy-efficient in this SHNO constructed by the [Ni/Co]$_n$(1.7 nm)/Pt(4 nm) multilayer. Figures ~\ref{fig2}(a) and ~\ref{fig2}(b) show two representations of the generated microwave spectra obtained at $I$ = 5.5 mA and 6.5 mA, $\theta$ = 120$^\circ$ and the external magnetic field ranging from 200 to 2000 Oe. The central peak frequency $f_{auto}$ of auto-oscillation [Fig.~\ref{fig2}(c)] can be extracted by fitting the generated spectral peak using the Lorentzian function [the solid curves in Figs.~\ref{fig2}(a) and \ref{fig2}(b)]. To obtain the magnetic properties of the [Co/Ni] multilayer and the relationship between the observed auto-oscillation mode and the uniform FMR mode, we measured the field-dependence of FMR of the device by using the ST-FMR technique~\cite{yanglp,wang2022}. Figure~\ref{fig2}(c) shows the field dependence of the FMR frequency $f_{FMR}$ and auto-oscillation frequencies $f_{auto}$ obtained at $I$ = 5.5 mA and 6.5 mA. The $f_{auto}$ is higher than $f_{FMR}$ at small fields $H \leq$ 1.6 kOe for a low driving current $I$ 5.5 mA. Since the auto-oscillation shows a significant redshift with the driving current $I$, $f_{auto}$ goes to be below $f_{FMR}$ for the large in-plane fields $H \geq$ 1.0 kOe at $I$ = 6.5 mA. The ST-FMR data are fitted by the Kittle formula $f = \gamma\sqrt{H(H+4\pi M_{eff})}$ with a fitting parameter $4\pi M_{eff}$ = 2.9 kOe [solid curve in Fig.~\ref{fig2}(c)]. The effective demagnetizing field can be expressed as the following form of $4\pi M_{eff} = 4\pi M_s - \frac{2K_u}{M_s}$, where $M_s$ is the saturation magnetization, $K_u$ is the uniaxial anisotropy coefficient. The PMA coefficient $K_u$ = 0.25 MJ/m$^3$ is determined from the FMR resonance frequency $vs.$ field dispersion curve and the saturation magnetization of the film.

Furthermore, we analyze the dependence of the generation spectra on the in-plane angle $\theta$ formed by the applied field relative to the direction of the current flow. Figure ~\ref{fig2}(d) shows that the auto-oscillation frequency substantially decreases when the angle approaches $\theta$ = 90$^\circ$. Based on the symmetry of SHE, the maximum excitation efficiency is reached when the spin polarization of spin currents generated by the Pt layer is antiparallel to the magnetization of the [Co/Ni]$_n$ multilayer, corresponding $\theta$ = 90$^\circ$~\cite{wang2022}. Therefore, the observed frequency decrease toward $\theta$ = 90$^\circ$ is consistent with that the excited spin-wave mode has a strong frequency redshift with increasing excitation current or $\theta$-induced excitation efficiency. It should be noted that the microwave spectral peaks vanish at $\theta$ approaching 90$^\circ$. The reason is that the microwave signal is generated due to the AMR of [Co/Ni], which has the sinusoidal dependence on the orientation of $\mathbf{H}$ (or $\mathbf{M}$) with a period of 180$^\circ$~\cite{wang2022,wangPRA}, magnetization oscillation cannot generate a signal at the fundamental harmonic of oscillation at angles close to 90$^\circ$ corresponding $dR_{AMR}/d\theta$ = 0.

\begin{figure*}[!t]
\centering
\includegraphics[width=0.95\textwidth]{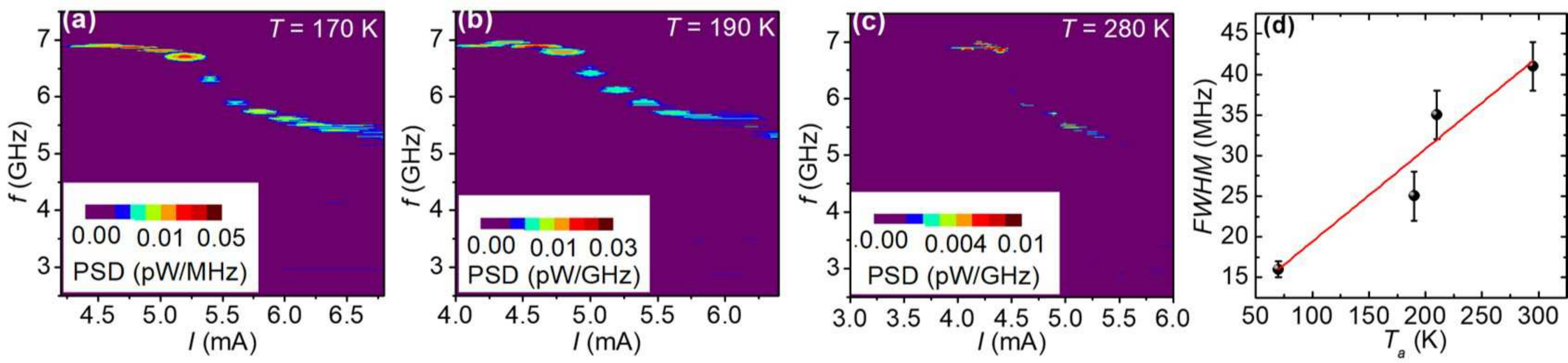}
\caption{Dependence of the microwave generation characteristics on current at different temperatures $T$, $H$ = 1.0 kOe, and $\theta = 120^\circ$. (a-c) Pseudocolor plots of the spectra obtained above $I_c$ increased in 0.2 mA steps, $T$ = 170 K (a), 190 K (b), 280 K (c). (d) The calculatedly actual temperature $T_a$ vs. the minimum linewidth $FWHM$ (symbols), corresponding to the highest intensity peak of PSD spectra, determined by fitting the spectra of (a-c) and Fig.~\ref{fig3}(i) with Lorentzian function. The solid line is the linear fit.}\label{fig4}
\end{figure*}

To further confirm the nonlinear frequency redshift of the auto-oscillation mode, the current-dependencies of the generated microwave spectra are measured at three representative fields $H$ = 100, 400, and 1000 Oe, $\theta$ = 120$^\circ$ and 6 K, as shown in Fig.~\ref{fig3}. At a very small field $H$ = 100 Oe [Fig.~\ref{fig3}(a)], a peak with a frequency higher than $f_{FMR}$ begins to appear in microwave spectra at the onset of current $I_{on}$ = 5.0 mA. Its frequency shows a near linear redshift with the excitation current [Fig.~\ref{fig3}(b)]. The linewidth shows a quasilinear decrease near above $I_{on}$ and reaches a minimum value of 50 MHz at $I$ = 7.2 mA, corresponding to the maximum peak power spectral density (PSD) [Fig.~\ref{fig3}(c)-\ref{fig3}(d)]. This behavior is consistent with the theoretical model of spin-torque nano-oscillation in which the thermal linewidth will decrease with increasing oscillation power~\cite{nonlinear}. Above 7.2 mA, the spectral peak begins to broaden, and its magnitude decreases with a more significant frequency redshift. For the medium field $H$ = 400 Oe [Fig.~\ref{fig3}(e)], the oscillation peak shows a noticeable blueshift, a linear decrease of the linewidth and a rapid increase of power with increasing current at $I > I_{on}$ = 4.5 mA [Fig.~\ref{fig3}(f)-\ref{fig3}(h)]. The linewidth decreases to a minimum value of 10 MHz at the same current $I$ = 6.2 mA as the maximum peak PSD and the maximum frequency, and then increases with continued increasing current, accompanied by a significant redshift and a decrease of the peak PSD. At the relatively large field $H$ = 1000 Oe [Fig.~\ref{fig3}(i)], the dependence of oscillation peak on the excitation current exhibits similar overall behavior with the medium field 400 Oe except for having a frequency more close to $f_{FMR}$ at the small currents and a more significant redshift behavior at the larger currents [Fig.~\ref{fig3}(j)-\ref{fig3}(l)]. The linewidth rapidly increases from its minimum value of 10 MHz at 5.5 mA and exhibits a peak at 6 mA, also correlated with the onset of a large frequency redshift~\cite{kim}. The increase of the onset current $I_{onset}$ and the minimum linewidth at small fields is likely correlated to the oscillation frequency close to the linear spin-wave spectrum, resulting in large damping due to their overlap or the magnetic anisotropy field-induced inhomogeneity of magnetic properties at low fields.

\subsection{Temperature effect on current-driven dynamical mode}

To explore the thermal effects on the spectral coherence of the generated microwave signals in this SHNO with a moderate interfacial magnetic anisotropy, we repeat the generated microwave spectra with a large field $H$ = 1.0 kOe at different selected temperatures.  Figure~\ref{fig4}(a)-\ref{fig4}(c)show the microwave-generation spectra acquired at additional three experimental temperatures $T$ = 170 K, 190 K and 280 K, similarly to the behaviors at $T$ = 6 K discussed above[Fig.~\ref{fig3}(i)]. We note that the actual temperature of the active device area is higher than the experimental temperatures due to current-induced Joule heating. Similar to our previous works, we can quantitatively obtain the actual device temperature $T_a$ by directly comparing $R(I)$ and $R(T)$ curves at each experimental temperature~\cite{liurh} or the COMSOL MULTIPHYSICS simulation of Joule heating of the device~\cite{prb2019}. In addition, as discussed above [Fig.~\ref{fig3}], the nonlinearity can dramatically reduce the oscillation coherence and significantly broaden the linewidth of spectra at the large current-induced redshift region~\cite{nonlinear,kim}. To avoid these anomalous contributions, we analyze the minimum value of the linewidth at the current corresponding to the maximum peak PSD and the highest frequency. Figure~\ref{fig4}(d) shows that the minimum linewidth of the oscillation mode approximately follows a linear temperature dependence. This linear dependence is consistent with the previously reported traditional spin-transfer-torque nano-oscillators~\cite{linear,thermaleffect,Fert-Tem-Vor} and SHNOs with a PMA FM layer~\cite{liupra2020}, but is contrasted with the thermal effects in the planar-nanogap SHNOs with in-plane magnetized Py without PMA, where the linewidth shows an exponential dependence on temperature due to thermally activated transitions between the primary bullet and secondary edge modes~\cite{prb2019,temline}. Previous nonlinear theory of spin-torque nano-oscillations with a single-mode indicated that the thermal noise could result in a linear broadening of linewidth with temperature~\cite{tiberkevich,silvather}, well consistent with the demonstrated single-mode nature of the magnetization dynamics in [Co/Ni]-based SHNOs with PMA.

\begin{figure*}[!t]
\centering
\includegraphics[width=0.95\textwidth]{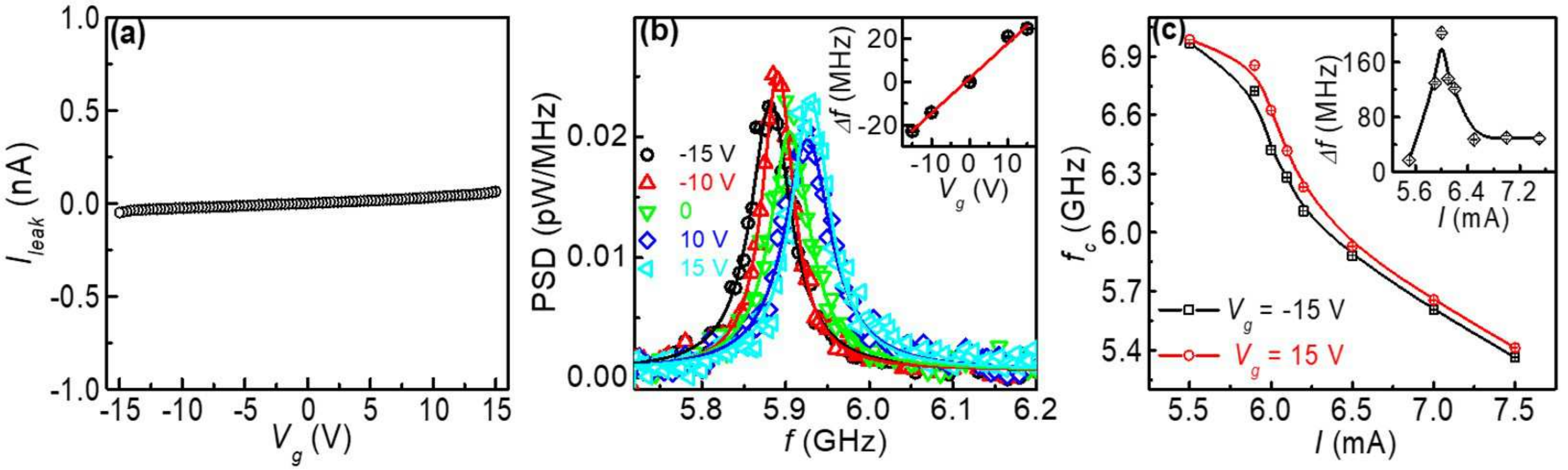}
\caption{Effects of voltage gating on the generated microwave spectra of SHNO at 6 K. (a) Gate leakage current $I_{leak}$ vs. gate voltage $V_g$ for a gated SHNO device. (b) Symbols: Power spectral density (PSD) of generation spectra at the labeled values of the gate voltage $V_g$ ranging from -15 to 15 V, at $H$ = 1000 Oe and $I$ = 6.5 mA. The curves are the results of fitting by the Lorentzian function. Inset: dependence of the central frequency shift $\Delta f(V_g) = f_c(V_g) - f_c(0)$ of the spectral peak on the gate voltage (symbols), and the linear fit of the data (line). (c) Dependence of the central generation frequency $f_c$ on current $I$ at $V_g$ = -15 V (squares) and 15 V (circles). The solid lines are given as guides to the eye. Inset: dependence of the central frequency shift $\Delta f(V_g) = f_c(15 V) - f_c(-15 V)$ between $V_g$ = 15 V and -15 V (left vertical axis) and the differential $df/dI$ (right vertical axis) on excited current $I$.} \label{fig5}
\end{figure*}

\subsection{Electric-field effect on current-driven dynamical mode}

Besides current-induced SOTs, voltage-controlled magnetic anisotropy (VCMA) offers an alternative approach to manipulate the damping constant and direction of magnetization~\cite{matsukura,weisheit,shiota,liuprb,liu2017,fulara2}. Therefore, we further investigate current- and voltage-based collaborative control of nonlinear magnetization oscillations and spin-waves in three-terminal SHNOs. Figure~\ref{fig5}(a) shows the dependence of the leakage current $I_{leak}$ between the Pt/[Co/Ni] layer and the Cu gate electrode on the voltage $V_g$ applied to the gate. The leakage does not exceed 0.1 nA at gate voltages of up to $\pm$ 15 V, indicating a high quality of the 30 nm thick BaTiO$_3$(30) insulator characterized by the breakdown electric field of more than 5 MV/cm. We analyze the dependence of the oscillation characteristics on the gate voltage $V_g$ at field $H$ = 1.0 kOe, in which the SHNO shows a significant redshift and a large generation power, as discussed above. Figure~\ref{fig5}(b) shows the power spectral density of the oscillation spectra at $I$ = 6.5 mA for the bias voltage $V_g$ ranging from -15 to 15 V. The shift $\Delta f = f_c(V_g) - f_c(0)$ of the central oscillation frequency exhibits a linear dependence on $V_g$ with the slope of 1.5 MHz/V at $I$ = 6.5 mA, as shown in the inset of Fig.~\ref{fig5}(b). This positive slope is contrary to the negative trend in the previously reported nano-constriction W(5)/CoFeB(1.7)-based SHNO~\cite{fulara2}. In that case, the SHNO has a large PMA value $K_u \simeq $ 0.6 MJ/m$^3$, and exhibits a considerable frequency blueshift with the driven current at a large external magnetic field with an out-of-plane oblique angle of 60$^{\circ}$. The negative trend of the voltage-controlled modulation of auto-oscillation frequency is caused by the overall effect of two opposite contributions. For instance, from the FMR Kittel formula, the linear increase in the interfacial PMA coefficient $K_u$ with negative gate voltage will result in a significant increase in the oscillating frequency; While the increase of the threshold current with negative gate voltage and blueshift with the driving current leads to an effective decrease of frequency. Therefore, to gain insight into the positive voltage modulation of frequency for our case, we further analyze the current dependence of the oscillation characteristics at $\pm V_g$.

Figure~\ref{fig5}(c) shows the current-dependence of the central oscillation frequency acquired at $V_g$ = -15 and 15 V. One can easily see that the effect of gating on the oscillation frequency can be described as being mostly a driving current shift, which is caused by the voltage-controlled change in the effective damping constant or/and the interfacial Rashba dampinglike torque efficiency in prior W/CoFeB-based nano-constriction SHNO with a large PMA~\cite{fulara2} and Py/Pt-based nano-gap SHNO without PMA~\cite{liu2017}. However, it should be noted that the sign of voltage-controlled modulation of the threshold current in the studied BTO/[Co/Ni]/Co/Pt is opposite to those prior two SHNOs. The reason may be related to the different types of the excited spin-waves mode (local bullet and propagating modes) or the different electronic band structures at different FM/insulators. The frequency difference $\Delta f = f_c(V_g = 15 V) - f_c(V_g = -15 V)$ vs. the driving current $I$ curve [left vertical axis of inset in Fig.~\ref{fig5}(c)] almost overlaps with the current-dependent redshift rate $df/dI$ [right vertical axis of inset in Fig.~\ref{fig5}(c)], which further confirms that the argument of the observed voltage-modulated frequency mainly coming from the excitation current shift. Directly comparing these two curves, we obtain the gating voltage-modulated excitation current shift of $\pm$ 0.05 mA at $V_g = \pm$ 15 V. The maximum of 200 MHz voltage-controlled frequency tunability is achieved in this three-terminer nano-gap SHNO. Our results demonstrate that a high speed and large gating-tunability of oscillation frequency can be achieved by combining current-dependent redshift and voltage-controlled magnetic anisotropy and interfacial Rashba dampinglike torque or nonlinear damping.

\subsection{Micromagnetic simulations and mechanism for achieving single-mode oscillation}

\begin{figure}[htbp]
\centering
\includegraphics[width=0.45\textwidth]{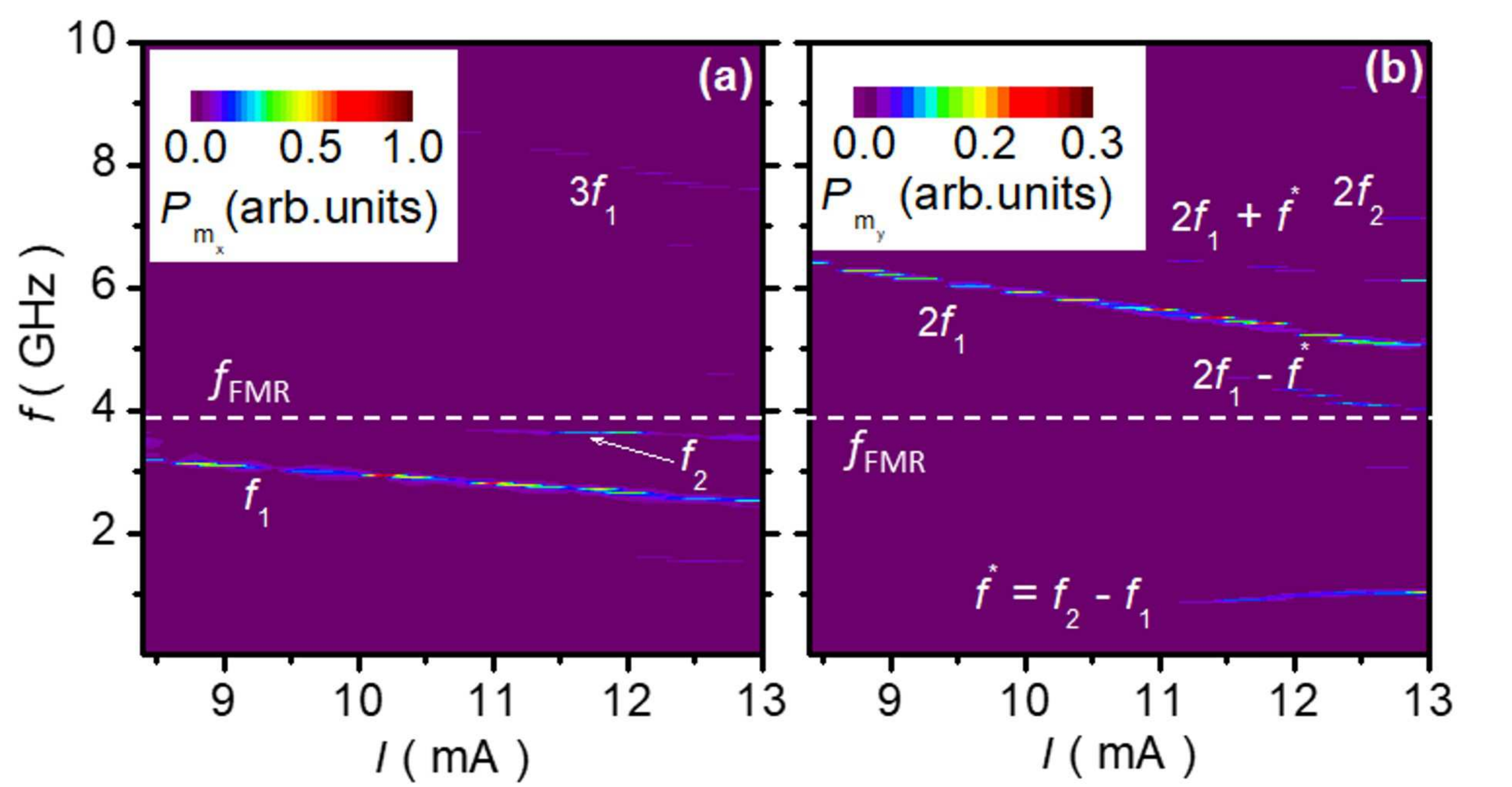}
\caption{Pseudocolor maps of the dependence of power spectra on current for an SHNO with $K_u$ = 0 and $H$ = 200 Oe. The power spectrum was obtained by performing the fast Fourier transform (FFT) of temporal in-plane components $m_x$ (a) and $m_y$ (b) of magnetization. The in-plane magnetic field is along the y-axis. The different peaks corresponding to the distinct dynamical modes were labeled by the primary mode $f_1$, its second- and third-harmonics 2$f_1$ and 3$f_1$, the secondary mode $f_2$, its second-harmonics 2$f_2$ and the intermodes $f^* = f_2 -f_1$, 2$f_1 \pm f^*$. The FMR frequency $f_{FMR}$ was marked by the dashed line.}\label{fig6}
\end{figure}

\begin{figure}[htbp]
\centering
\includegraphics[width=0.45\textwidth]{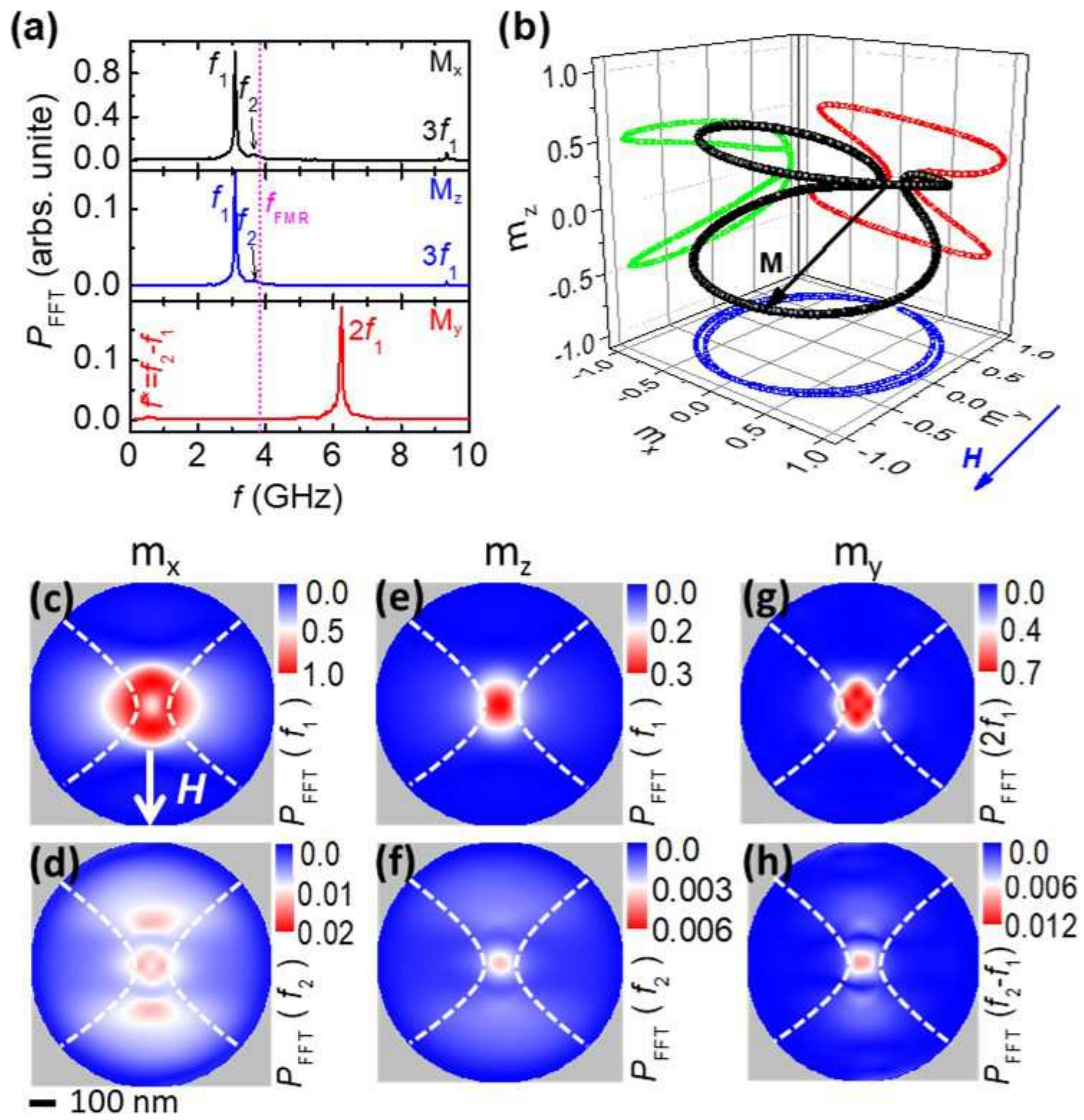}
\caption{(a) Representative FFT power spectra of three magnetization components $m_x$, $m_z$ and $m_y$ for an SHNO with $K_u$ = 0 obtained at $H$ = 200 Oe and a small current $I$ = 9 mA. The vertical dashed line presents its FMR frequency $f_{FMR}$. (b) 3D-plot of the trajectory of magnetization $\mathbf{M}$ (represented by the black arrow) located at the central nanogap region. (c)-(h) Spatial power maps (normalized by the maximum of $f_1(m_x)$) of $m_x$ (c, d) and $m_z$ (e, f) at frequencies $f_1$ = 3.10 GHz and $f_2$ = 3.67 GHz, and $m_y$ (g, h) at second-harmonic 2$f_1$ = 6.2 GHz and intermode $f^* = f_2 - f_1$ = 0.57 GHz, respectively. Dashed lines show the contours of two top Au electrodes. The bold arrow indicates the direction of the magnetic field $H$.}\label{fig7}
\end{figure}

\begin{figure}[htbp]
\centering
\includegraphics[width=0.45\textwidth]{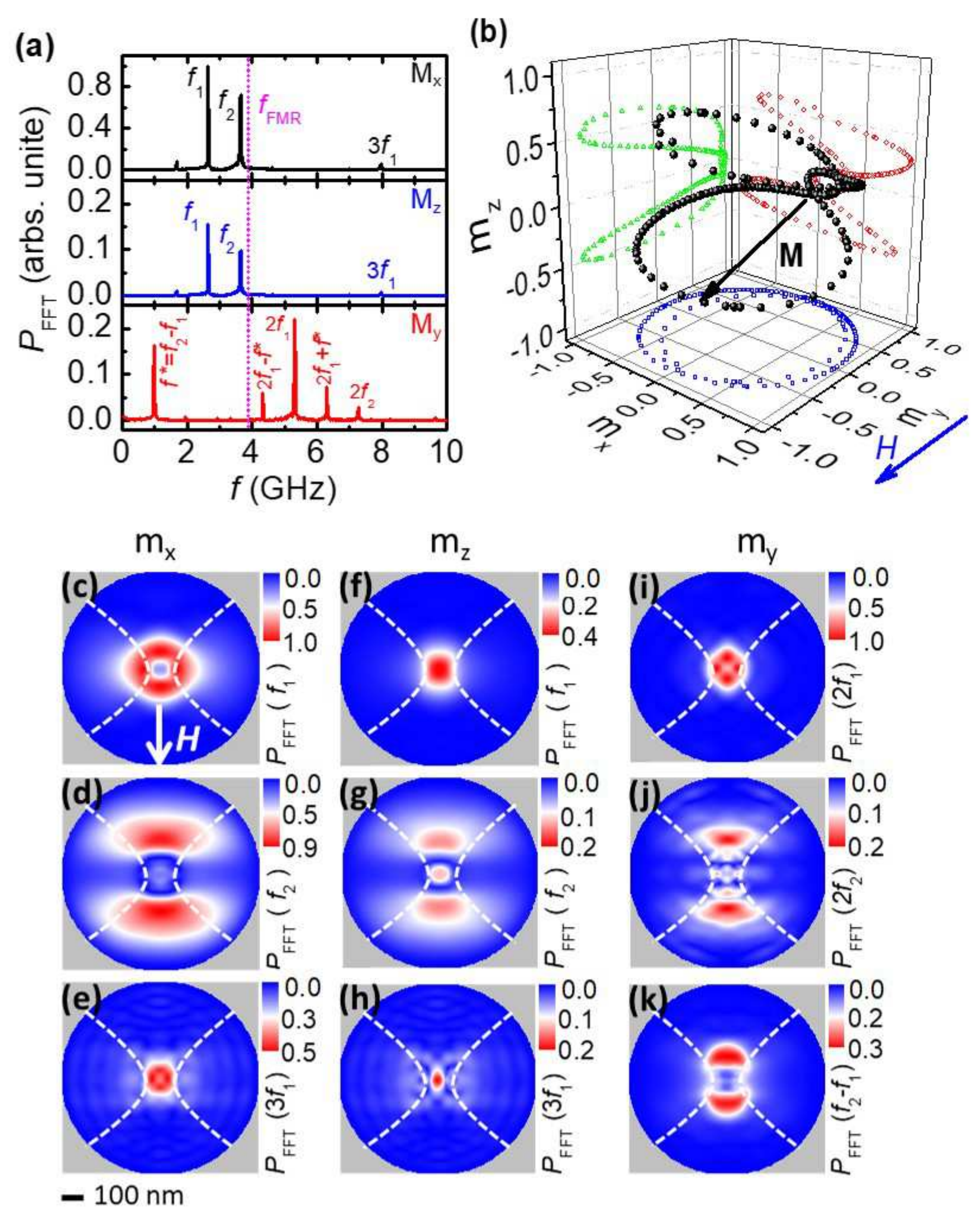}
\caption{(a) Typical FFT power spectra of $m_x$, $m_z$ and $m_y$ for an SHNO with $K_u$ = 0 obtained at $H$ = 200 Oe and a large current $I$ = 12 mA. The vertical dashed line presents the FMR frequency $f_{FMR}$. (b) 3D-plot of the trajectory of magnetization $\mathbf{M}$ (represented by the black arrow) located at the central nanogap region. (c)-(k) Spatial power maps (normalized by the maximum of $f_1(m_x)$) of $m_x$ (c-e) and $m_z$ (f-h) at frequencies $f_1$, $f_2$, 3$f_1$ and $m_y$ (i-k) at second-harmonics 2$f_1$, 2$f_2$ and intermode $f^* = f_2 - f_1$, respectively. The blue arrow and dashed lines show the applied field direction and the contours of the electrodes, respectively.}\label{fig8}
\end{figure}

To gain a physical understanding of the single dynamic mode observed experimentally in our SHNO with a moderate PMA, we perform micromagnetic simulations using the OOMMF software\cite{oommf}. The simulated volume is a circular disk with a diameter of 1 $\mu$m and a thickness of 2 nm, which is divided into 5 $\times$ 5 $\times$ 2 nm$^3$ cells. The following material parameters are used in the simulations: exchange stiffness $A$ = 10 pJ/m, saturation magnetization $M_s$ = 760 kA/m, Gilbert damping constant $\alpha$ = 0.03, effective STT efficiency $P$ = 0.07, and three typical PMA constants $K_u$ = 0, 0.2 and 0.35 MJ/m$^3$. To more precisely simulate a real SHNO, we perform the micromagnetic simulation using the actual spin current and Oersted field distributions, which are numerically calculated with the COMSOL MULTIPHYSICS package\cite{liu2019}. All micromagnetic simulations are done at $T$ = 0 and the local Joule heating effect is neglected. To conveniently illustrate the trajectory of magnetization $\mathbf{M}$, we select the magnetic field $H$ perpendicular to the current $I$ ($\theta$ = 90$^o$).

To directly compare with the previous experimental and simulation results of nano-gap SHNO without PMA, we first perform the dependence of spectra on the excitation current $I$. The calculated power spectrum is obtained by performing the fast Fourier transform (FFT) of the time series of the in-plane $m_x$ [Fig.~\ref{fig6}(a)] or out-of-plane magnetization components $m_z$ [Fig.~\ref{fig6}(b)]. Figure~\ref{fig6}(a) shows the dependence of the power spectra on $I$ at $H$ = 200 Oe characterized by the primary low-frequency bullet mode $f_1$ with a frequency far below $f_{FMR}$ and a significant redshift appeared first, and followed by the coexistence with the secondary high-frequency mode $f_2$ at large currents. These results are consistent with the previous experimental observations and simulation results at $\theta$ = 120$^o$ ~\cite{liurh,prb2019} and 90$^o$~\cite{demidov,ulrichs}. In addition, for in-plane SHNO with an effective easy-plane shape anisotropy, the precessing magnetization vector $\mathbf{M}$ creases dynamical demagnetizing field antiparallel to the out-of-plane component $m_z$ of magnetization, and forces $\mathbf{M}$ to do elliptical precession with the short axis normal to the film plane under spin torque at an in-plane magnetic field, as shown in Fig.~\ref{fig7}(b). Besides the fundamental frequency, the elliptical precession also exhibits by the oscillation of the component of $\mathbf{M}$ parallel to the applied field $m_y$ at twice the frequency of precession, which is well consistent with the FFT spectrum of $m_y$ in Fig.~\ref{fig6}(b). In addition to the second-harmonic of the primary bullet mode $f_1$ and secondary mode $f_2$, the strong intermodes $f^* = f_2 - f_1$, 2$f_1 \pm f^*$ are also observed in the calculated power spectrum corresponding to $m_y$ [Fig.~\ref{fig6}(b)], indicating that there exist the strong nonlinear coupling between $f_1$ and $f_2$. It is also consistent with recent Brillouin light scattering (BLS) spectra experiment and simulations~\cite{nonlineardamping}, which revealed that the nonlinear coupling in this spin Hall nano-device with the extended magnetic film is determined by the ellipticity of magnetization precession.

\begin{figure}[htbp]
\centering
\includegraphics[width=0.45\textwidth]{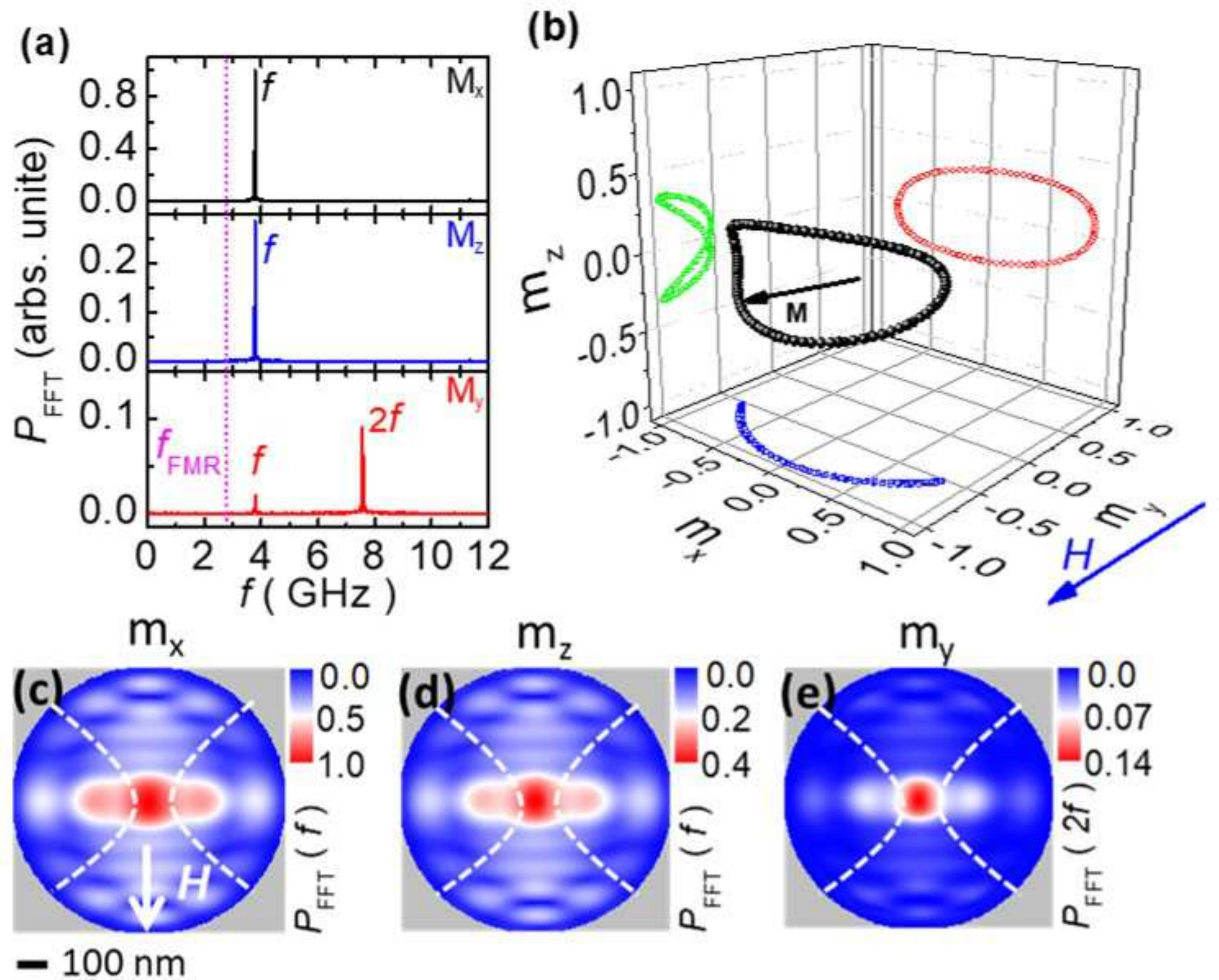}
\caption{(a) Typical FFT power spectra of $m_x$, $m_z$ and $m_y$ for an SHNO with a moderate $K_u$ = 0.2 MJ/m$^3$ obtained at $H$ = 200 Oe and $I$ = 3.5 mA. The vertical dashed line presents its FMR frequency $f_{FMR}$. (b) 3D-plot of the trajectory of magnetization $\mathbf{M}$ (represented by the black arrow) located at the central nanogap region. (c)-(e) Spatial power maps (normalized by the maximum of $f_1(m_x)$) of $m_x$ (c) and $m_z$ (d) at the fundamental frequency $f$ and $m_y$ (e) at its second-harmonic 2$f$, respectively. Dashed lines and bold arrow show the contours of the electrodes and the applied field direction, respectively.}\label{fig9}
\end{figure}

To inspect the excitation mechanism of the secondary mode $f_2$ and its relation to the primary bullet mode, we need to analyze the calculated spectrum, precession trajectory of $\mathbf{M}$, and their spatial profiles comprehensively. At a small current $I$ = 9 mA, the power spectrum was dominated by the low-frequency mode $f_1$ for $m_x$ and $m_z$, but the second-harmonic 2$f_1$ for $m_y$ [Fig.~\ref{fig7}(a)] due to the easy-plane anisotropy-induced elliptical precession [Fig.~\ref{fig7}(b)]. The power intensity of the high-frequency mode $f_2$ is less than 2\% of $f_1$. The spatial power maps with three components ($m_{x,y,z}$) corresponding to the two modes and their intermode $f_1-f_2$ are shown in Figs.~\ref{fig7}(c)-~\ref{fig7}(h). We note that the spatial power maps of $f_2$ [Figs.~\ref{fig7}(d) and ~\ref{fig7}(f)] include a certain background signal of $f_1$ due to its tiny power intensity compared to $f_1$ and a small frequency difference between $f_1$ and $f_2$.

To get some more insight into the excitation mechanism of the secondary mode, we further analyze the auto-oscillation dynamical characteristics obtained at a large current $I$ = 12 mA, as shown in Fig.~\ref{fig8}. The high-frequency mode $f_2$ is significantly enhanced and has a power intensity comparable to that of the bullet mode $f_1$. Similar to $I$ = 9 mA [Fig.~\ref{fig8}], the primary bullet mode $f_1$ is localized in the center nano-gap region of the device. In contrast, the secondary mode $f_2$ is localized much weaker compared to $f_1$ and exhibits two maxima located at a distance of about 150 nm from the center of the gap in two opposite directions collinear with the field. We note that $m_x$ exhibits a larger spatial distribution than $m_z$, which may be related to the in-plane magnetization and large oscillation amplitude of $m_x$. These characteristics are consistent with the prior micro-focused BLS measurements and simulations~\cite{demidov,ulrichs,prb2019}. Previous simulations infer that the secondary edge mode is stabilized by two effective potential wells created by the dipole field of the primary bullet mode~\cite{ulrichs,prb2019}. However, in the outer region of the nanogap, the spin current density $J_s$ is too low to directly excite or maintain the high-frequency edge mode $f_2$ because more than 80\% $J_s$ is localized in the center nanogap~\cite{prb2019}. There must have an intermediary to transfer energy from the center nanogap region for compensating energy dissipation of the outer edge spin-wave mode. Numerous nonlinear theories and experiments have revealed that the nonlinear spin-wave coupling can enable energy transfer between different modes resulting in mode coexistence, transition, hopping and chaos phenomena~\cite{liu2015,prb2019,nonlinear,nonlineardamping,Dumas,MC1,zhang,MC2,MC3}. Therefore, the formation of the secondary mode coexisting with the center bullet mode is likely attributed to the nonlinear coupling-induced energy transfer mechanism~\cite{nonlineardamping,MC1,MC2} and localized by the effective potential well generated by the spatially inhomogeneous dipole field raised from the center bullet mode~\cite{ulrichs,prb2019}. The latter determines the spatial location of the secondary mode $f_2$.

\begin{figure*}[!t]
\centering
\includegraphics[width=0.95\textwidth]{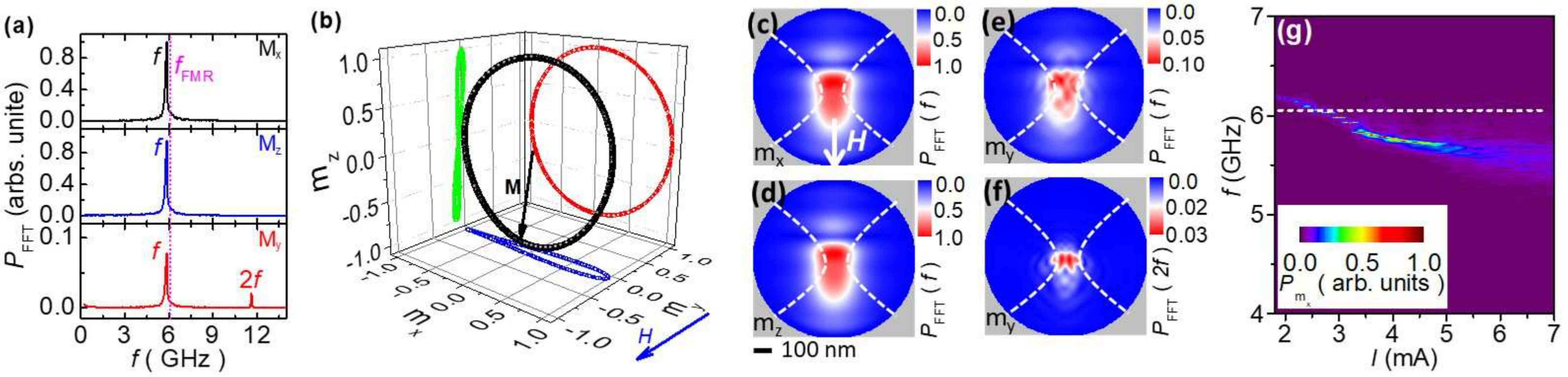}
\caption{(a) Typical FFT power spectra of $m_x$, $m_z$ and $m_y$ for an SHNO with a lager $K_u$ = 0.35 MJ/m$^3$ obtained at $H$ = 2000 Oe and $I$ = 3 mA. The vertical dashed line presents its FMR frequency $f_{FMR}$. (b) 3D-plot of the trajectory of magnetization $\mathbf{M}$ (represented by the black arrow) located at the central nanogap region. (c)-(f) Spatial power maps (normalized by the maximum of $f_1(m_x)$) of $m_x$ (c) and $m_z$ (d) at the fundamental frequency $f$ and $m_y$ (e, f) at $f$ and its second-harmonic 2$f$, respectively. Dashed lines and bold arrow show the contours of the electrodes and the applied field direction, respectively. (g) Pseudocolor map of the dependence of power spectra on current at $H$ = 2.0 kOe}\label{fig10}
\end{figure*}

How to understand the nonlinear spin-wave mode coupling? As the discussion above [Fig.~\ref{fig6}], the oscillation frequency of $m_y$ is twice that of the primary bullet mode $f_1$ of $m_x$ and $m_z$ due to the elliptical precession [Fig.~\ref{fig7}(b) and Fig.~\ref{fig8}(b)]. The frequencies of the second (2$f_1$) and third (3$f_1$) harmonics are above $f_{FMR}$, within the linear spectrum of propagating spin waves. Therefore, the central local magnetization dynamics at these frequencies, associated with the bullet mode precession, can be expected to couple to propagating spin waves at the corresponding frequencies, and acts as a parametric pump~\cite{nonlineardamping,parameter1,parameter2} that drives energy transfer from the primary bullet mode $f_1$ into the secondary edge mode $f_2$, resulting in nonlinear damping and frequency decrease of the former. Indeed, the maps of these harmonics show intensity modulations consistent with spin-wave radiation from the central bullet regime. The observed strong intermode $f^* = f_2 - f_1$ at large currents also confirms nonlinear coupling between the nonlinear bullet $f_1$ and the high-frequency edge modes $f_2$.

As far as we know, the effective PMA field can counteract the demagnetizing field. Therefore, one can expect to diminish the ellipticity of magnetization precession with the help of the PMA field, and suppress the nonlinear damping of the dominated spin wave mode discussed above. Furthermore, the secondary edge mode supported by a nonlinear coupling-induced energy transfer mechanism is expected to become suppressed in our [Ni/Co]/Pt-based SHNO with PMA compensating shape anisotropy. Additionally, the PMA field can also reduce the negative nonlinearity coefficient $\aleph$ for the in-plane magnetized thin film, and drive nonlinearly localized mode to a propagating spin wave mode through elevating the oscillation frequency~\cite{Fulara,fulara2,nonlinear,Dumas,droplet}. To verify this argument above, we bring in the different PMA constants $K_u$ in our simulations of SHNO. First, we choose a relatively small 0.2 MJ/m$^3$, less than the demagnetizing field. In the same way as before, the calculated spectrum, precession trajectory and their spatial profiles are analyzed, as illustrated in Fig.~\ref{fig9}. There are several differences from the case of $K_u$ = 0. First, the power intensity of magnetization component $m_y$ at twice frequency 2$f$ was reduced by more a half compared to the case without PMA [Fig.~\ref{fig7}(a) and Fig.~\ref{fig8}(a)], consistent with the ellipticity decrease of magnetization precession in trajectory chart [Fig.~\ref{fig9}(b)]. Second, the stable auto-oscillation can be achieved at a small driving current $I$ = 3.5 mA, far below 8 mA of $K_u$ = 0 [Fig.~\ref{fig6}], consistent with the diminishment of nonlinear damping discussed above. Third, in contrast to two-mode coexistence, a single dynamical mode is only observed at a small current 3.5 mA, and its frequency is above $f_{FMR}$, consistent with the character of a propagating spin wave mode with intensity modulations in profile mapping [Fig.~\ref{fig9}(c) - ~\ref{fig9}(e)]. We note that, in this case of $K_u$ = 0.2 MJ/m$^3$, the auto-oscillation still exhibits a redshift with a negative $\aleph$ and elliptical precession. Consequently, the frequency of the propagating mode can be driven by a large current to below $f_{FMR}$ and becomes a localized mode or two modes coexistence at large currents and high in-plane fields.

We further analyze the case with $K_u$ = 0.35 MJ/m$^3$ equal to the demagnetizing field. Because the PMA and demagnetizing fields compensate each other, the magnetization precession becomes circular consistent with trajectory results obtained by simulation, shown as in Fig.~\ref{fig10}(b). In contrast to the small $K_u$ [Figs.~\ref{fig7}(a), ~\ref{fig8}(a) and ~\ref{fig9}(a)], the power spectrum associated with $m_y$ becomes be dominated by the fundamental frequency [Figs.~\ref{fig10}(a)]. As follows from the discussions above, nonlinear damping due to nonlinear coupling-induced energy transfer from the central dominant mode is further minimized, which is supported by the calculated auto-oscillation exhibiting a well-defined single-mode and more low threshold current. In addition to a high oscillation frequency above $f_{FMR}$[Fig.~\ref{fig10}(a)], the spatial power maps corresponding to three magnetization components show a large area of asymmetric elongated spatial profile with certain intensity modulations and the direction of elongation (propagating) along the in-plane magnetic field [Fig.~\ref{fig10}(c) - \ref{fig10}(f)]. These characteristics are well consistent with the linear propagating spin wave mode. Figure~\ref{fig10}(g) shows that the auto-oscillation shows a noticeable redshift with increasing the excitation current and becomes a local spin wave mode at the large current due to the negative nonlinearity coefficient and local Oersted field. We note that the out-of-plane oblique magnetic field is also expected to compensate for the shape anisotropy, modulate nonlinearity coefficient $\aleph$ and nonlinear damping, and achieve a single-dynamical mode with high coherence in SHNOs without PMA by effectively suppressing nonlinear mode coupling induced secondary edge-mode.

\section{Conclusions}

To summarize, in a three-terminal spin Hall nano-oscillator based on BaTiO$_3$/[Co/Ni]/Pt trilayers with a moderate interfacial PMA, we achieve good coherent single-mode dynamic with a current-modulated oscillation frequency rate of 15\%/mA at $f_{auto}$ = 5.5 GHz (or 0.85 GHz/mA) and a voltage-controlled tunability of frequency $\sim$ 200 MHz. The current-modulated frequency is related to the intrinsic nonlinearity of the nano-oscillator and the nonlinear damping due to nonlinear coupling between the fundamental auto-oscillation mode and elliptical precession causing its higher order harmonics, within the linear spectrum of propagating spin waves. In comparison, the voltage-control mainly originates from gating voltage-induced current threshold shift due to the effects of electrostatic gating on the interfacial Rashba dampinglike torque and/or the effective damping constant. Furthermore, the simulations of the PMA-dependent auto-oscillation demonstrate that the secondary high-frequency mode, usually observed in in-plane magnetized nano-gap SHNO without PMA, is attributed to the combination of the nonlinear mode coupling and the spatially inhomogeneous dipole field generated by the center bullet mode. The nonlinear mode coupling, determined by the ellipticity of the magnetization precession, can be diminished by utilizing the effective PMA field to compensate for the demagnetizing field induced by shape anisotropy. Additionally, the PMA field also can drive nonlinearly self-localized bullet mode to a quasi-linear propagating mode by suppressing the negative nonlinearity coefficient $\aleph$ and nonlinear damping. The simulation results support our experimentally observed coherent single-mode spin-wave in an SHNO with a moderate interfacial PMA. Closely associated with the modulation of nonlinear damping and mode coupling, the energy-efficient gate-voltage and current control of the oscillator demonstrated here can significantly facilitate the development of SHNO-based on-chip microscale microwave generators and neuromorphic computing.

\space
\textbf{Acknowledgements}\\
This work was supported by the National Natural Science Foundation of China (Grant Nos. 12074178, 12004171 and 11874135), the Applied Basic Research Programs of Science and Technology Commission Foundation of Jiangsu Province, China (Grant No. BK20200309), Key Research and Development Program of Zhejiang Province (Grant No.2021C01039), the Open Research Fund of Jiangsu Provincial Key Laboratory for Nanotechnology, and Postgraduate Research \& Practice Innovation Project of Jiangsu Province (Grant No. KYCX210699).

\end{document}